\pdfoutput=1

\documentclass[11pt,twoside,a4paper,cmspaper,final,collab]{cms-tdr}
\def\svnVersion{177386}\def\cmsCernNo{2012-175}\def\cmsCernDate{\today}\def\cmsMessage{Submitted to Physics Letters B}

\begin{document}\cmsNoteHeader{TOP-11-030}

\hyphenation{had-ron-i-za-tion}
\hyphenation{cal-or-i-me-ter}
\hyphenation{de-vices}

\RCS$Revision: 145318 $
\RCS$HeadURL: svn+ssh://svn.cern.ch/reps/tdr2/papers/TOP-11-030/trunk/TOP-11-030.tex $
\RCS$Id: TOP-11-030.tex 145318 2012-08-29 18:29:40Z alverson $
\newlength\cmsFigWidth
\ifthenelse{\boolean{cms@external}}{\setlength\cmsFigWidth{0.85\columnwidth}}{\setlength\cmsFigWidth{0.4\textwidth}}
\ifthenelse{\boolean{cms@external}}{\providecommand{\cmsLeft}{top}}{\providecommand{\cmsLeft}{left}}
\ifthenelse{\boolean{cms@external}}{\providecommand{\cmsRight}{bottom}}{\providecommand{\cmsRight}{right}}
\cmsNoteHeader{TOP-11-030} 
\title{Inclusive and differential measurements of the \texorpdfstring{$\ttbar$}{t t-bar} charge asymmetry in proton-proton collisions at \texorpdfstring{$\sqrt{s} = 7\TeV$}{sqrt(s) = 7 TeV}}

\date{\today}

\abstract{The \ttbar charge asymmetry is measured in events containing a charged lepton (electron or muon) and at least four jets, one of which is identified as originating from b-quark hadronization. The analyzed dataset corresponds to an integrated luminosity of 5.0\fbinv collected with the CMS detector at the LHC. An inclusive and three differential measurements of the \ttbar charge asymmetry as a function of rapidity, transverse momentum, and invariant mass of the \ttbar system  are presented. The measured inclusive \ttbar charge asymmetry is $A_{C} = 0.004 \pm 0.010 ~\text{(stat.)}\,\pm 0.011 ~\text{(syst.)}$. This result and the three differential measurements are consistent with zero asymmetry as well as with the predictions of the standard model.}

\hypersetup{%
pdfauthor={CMS Collaboration},%
pdftitle={Inclusive and differential measurements of the t t-bar charge asymmetry in proton-proton collisions at 7 TeV},%
pdfsubject={CMS},%
pdfkeywords={CMS, physics, top quark, asymmetry, BSM}}

\maketitle 

\section{Introduction}

The top quark offers an excellent opportunity to search for departures from the standard model (SM) as its large mass makes it unique among all quarks. A possible hint for new physics contributions showing up in the top-quark sector is the discrepancy of the measured \ttbar forward-backward asymmetry with SM expectations, reported by the CDF~\cite{:2011kc} and D0~\cite{PhysRevD.84.112005} Collaborations at the Tevatron. These discrepancies are of the order of two standard deviations and even more in certain phase space regions. They have generated a large number of theoretical explanations that attribute them to contributions from physics beyond the standard model (BSM). An overview of the variety of theoretical explanations can be found, e.g., in Ref.~\cite{AguilarSaavedra:2011hz} and references therein.

The production of \ttbar pairs at leading order (LO) is symmetric with respect to the exchange of the top quark and antiquark. At higher-order calculations, QCD radiative corrections to the $\cPq\cPaq \rightarrow \ttbar$ process induce an asymmetry in the differential distributions of top quarks and antiquarks. The interference between initial-state and final-state radiation (ISR and FSR) processes as well as the interference between the Born and box diagrams generate a correlation between the direction of the top-quark momentum and that of the incoming quark, while the direction of the top-antiquark momentum is related to that of the incoming antiquark~\cite{Kuhn:1998jr}. While these processes induce a forward-backward asymmetry ($A_{FB}$) at the Tevatron $\Pp\Pap$ collider, the charge-symmetric pp collisions at the Large Hadron Collider (LHC) result in a different effect. At the LHC, the larger average momentum fraction of the valence quarks leads to an excess of top quarks produced in the forward and backward directions, while the top antiquarks are produced more centrally. This makes the difference of the absolute values of the rapidities of top quark and antiquark, $\Delta |y| = |y_{\cPqt}| - |y_{\cPaqt}|$, a suitable observable to measure this \ttbar charge asymmetry. The rapidity is defined as $y = \frac{1}{2}\ln(\frac{E+p_{z}}{E-p_{z}})$, where $E$ denotes the particle energy and $p_{z}$ its momentum component along the beam direction. For a given sensitive variable, the charge asymmetry is defined as:

\begin{equation}
A_{C} = \frac{N^{+} - N^{-}}{N^{+} + N^{-}}~~,
\label{eq:ACDC}
\end{equation}
where $N^{+}$ and $N^{-}$ represent the number of events with positive and negative values in the sensitive variable, respectively.

In pp collisions at the LHC, \ttbar production is dominated by gg fusion processes, while at the Tevatron, \ttbar pairs are dominantly produced via $\cPq\cPaq\rightarrow \ttbar$. As described above, only the latter process results in different angular distributions of top quarks and antiquarks and thus the charge asymmetry at the LHC is expected to be considerably smaller than the forward-backward asymmetry at the Tevatron. The SM prediction at next-to-leading order (NLO) precision is $A^{\mathrm{theory}}_{C} = 0.0115\pm 0.0006$~\cite{Kuhn:2011ri}.

Recently, the Compact Muon Solenoid (CMS) and ATLAS Collaborations have published first measurements of the charge asymmetry at the LHC and found respectively $A_{C} = -0.013 \pm 0.028~(\text{stat.})\,^{+0.029}_{-0.031}~(\text{syst.})$~\cite{Chatrchyan:2011hk} and $A_{C} = -0.019 \pm 0.028~(\text{stat.})\,\pm 0.024~(\text{syst.})$~\cite{Aad:2012ug}, consistent with the SM prediction. Although these results do not have the precision to establish a non-zero asymmetry at the LHC, they seem to disfavor large positive deviations from the SM prediction as seen for $A_{FB}$ at the Tevatron.
The potential disagreement between the Tevatron and LHC results might be due to BSM contributions having different effects on the Tevatron forward-backward asymmetry and the LHC charge asymmetry~\cite{Zupan, AguilarSaavedra:2012va}. On the other hand it is possible that the anomalous $A_{FB}$ values determined by the Tevatron experiments are due to incomplete theoretical predictions or unaccounted for systematic uncertainties.

To shed light on this question, it is crucial to not only measure the inclusive asymmetry but to also measure $A_{C}$ as a function of suitable variables enhancing the \ttbar charge asymmetry in certain regions.

In this Letter, we report on updated and further developed measurements of $A_{C}$, adopting the event selection, background estimation, and reconstruction of the \ttbar system from Ref.~\cite{Chatrchyan:2011hk}. We present an inclusive measurement and three differential measurements of the \ttbar charge asymmetry as a function of the rapidity, transverse momentum, and invariant mass of the \ttbar system, using the full 2011 dataset. Each of these three variables is sensitive to a certain aspect of the \ttbar charge asymmetry.

The rapidity of the \ttbar system in the laboratory frame, $|y_{\ttbar}|$, is sensitive to the ratio of the contributions from the $\cPq\cPaq$ and gg initial states to \ttbar production. The charge-symmetric gluon fusion process is dominant in the central region, while \ttbar production through $\mathrm{q\bar{q}}$ annihilation mostly produces events with the \ttbar pair at larger rapidities, which implies an enhancement of the charge asymmetry with increasing $|y_{\ttbar}|$~\cite{Kuhn:2011ri}.

The transverse momentum of the \ttbar pair in the laboratory frame, $p_{\mathrm{T}}^{\ttbar}$, is sensitive to the ratio of the positive and negative contributions to the overall asymmetry. The interference between the Born and the box diagrams leads to a positive contribution, while the interference between ISR and FSR results in a negative contribution. The presence of additional hard radiation implies on average a higher transverse momentum of the \ttbar system. Consequently, in events with large values of $p_{\mathrm{T}}^{\ttbar}$, the negative contribution from the ISR-FSR interference is enhanced~\cite{Kuhn:2011ri}.

The charge asymmetry is expected to depend on the invariant mass of the \ttbar system, $m_{\ttbar}$, since the contribution of the $\cPq\cPaq$ initial state processes is enhanced for larger values of $m_{\ttbar}$. This observable is also sensitive to new physics contributions; potential new heavy particles could be exchanged between initial quarks and antiquarks and contribute to the \ttbar production (see e.g., Ref.~\cite{Rodrigo:2010gm} and references therein). The amplitudes associated with these new contributions would interfere with those of the SM processes, leading to an effect on the \ttbar charge asymmetry, which increases as a function of the invariant mass of the \ttbar system.

\section{The CMS detector}

The central feature of the CMS apparatus is a superconducting solenoid, of 6\unit{m} internal diameter, providing a magnetic field of 3.8\unit{T}. Within the field volume are a silicon pixel and strip tracker, a crystal electromagnetic calorimeter (ECAL) and a brass/scintillator hadron calorimeter. CMS uses a right-handed coordinate system, with the origin at the nominal interaction point, the $x$ axis pointing to the centre of the LHC ring, the $y$ axis pointing up (perpendicular to the LHC plane), and the $z$ axis along the counterclockwise beam direction. The polar angle $\theta$ is measured from the positive $z$ axis and the azimuthal angle $\phi$ is measured in the $x$-$y$ plane. The pseudorapidity is defined as $\eta=-\ln (\tan\theta/2)$. The inner tracker measures trajectories of charged particles  within the pseudorapidity range $|\eta| < 2.5$, while the calorimeters provide coverage up to $|\eta| = 3.0$. The ECAL has an energy resolution of 3\,\% or better for the range of electron energies relevant for this analysis. Muons are measured in the pseudorapidity range $|\eta| < 2.4$, with gas-ionization detectors embedded in the steel return yoke.  Matching the muons to the tracks measured in the silicon tracker results in a transverse momentum resolution between 1 and 5\,\%, for $\pt$ values up to 1\TeVc. Extensive forward calorimetry complements the coverage provided by the barrel and endcap detectors. A more detailed description of CMS can be found in Ref.~\cite{JINST}.

\section{Data and simulation}

The measurements reported in this Letter are based on data taken with the CMS detector at a centre-of-mass energy of $7\TeV$, corresponding to an integrated luminosity of 5.0\fbinv. To translate the distributions measured with reconstructed objects to distributions for the underlying quarks, samples of simulated events are used. Top-quark pair events are generated with two different generators, either with \MADGRAPH version~5.1.1~\cite{madgraph} or with the NLO generator \POWHEG~\cite{Alioli:2010xd}. For both samples the parton shower is simulated using \PYTHIA version 6.4.24~\cite{pythia}, and the MLM parton shower/matrix element matching~\cite{MLM} in case of \MADGRAPH. Also the $\cPqt$ and tW channels of electroweak production of single top quarks are simulated using \POWHEG. The production of electroweak vector bosons in association with jets (\PW$+$jets and \cPZ$+$jets) is simulated using the same combination of \MADGRAPH and \PYTHIA as for the \ttbar signal. All samples are generated using the \PYTHIA Z2 Monte Carlo tune~\cite{Z2} to model the underlying event. The simulations include additional proton-proton interactions (pileup) with the same frequency of occurrence as observed in the analyzed data.

\section{Event selection and estimation of background}

The analysis uses \ttbar events where one of the \PW~bosons from the decay of a top-quark pair subsequently decays into a muon or electron and the corresponding neutrino, and the other \PW~boson decays into a pair of quarks originating jets. We therefore select events containing one electron or muon and four or more jets, at least one of which is identified as originating from \cPqb-quark hadronization. For the reconstruction of electrons, muons, jets, and any imbalance in transverse momentum due to the undetected neutrino, $\ETmiss$, we use a particle-flow (PF) algorithm~\cite{PFT-09-001}. Electron (muon) candidates are required to have $p_{\mathrm{T}} > 30 (20)\,\GeVc$ and be within $|\eta| < 2.5 (2.1)$, while jets are required to have $p_{\mathrm{T}} > 30\,\GeVc$ and $|\eta| < 2.4$. More details on the selection criteria applied to the events can be found in Ref.~\cite{Chatrchyan:2011hk}.

In total, 57\,697 events are selected, 24\,705 events in the electron$+$jets channel and 32\,992 events in the muon$+$jets channel. About 20\% of these events are expected to come from background processes like \PW$+$jets and \cPZ$+$jets production, the production of single top quarks, and multijet production. For the estimation of the background contributions we make use of the discriminating power of $\ETmiss$ and of M3, the invariant mass of the three jets with the largest vectorially summed transverse momentum. For each of the two lepton channels, these two distributions are fitted with a binned maximum-likelihood fit. For the \ttbar signal and the \PW$+$jets, \cPZ$+$jets, and single-top-quark background processes, the respective simulated samples are used to model the shapes of the $\ETmiss$ and M3 distributions, while an approach based on data from sideband regions featuring non-isolated leptons is used for the multijet background. Gaussian rate constraints are introduced into the likelihood function for the \cPZ$+$jets and single-top-quark processes according to the respective NLO cross sections, while the rates of all other processes are free parameters of the fit. A more detailed description of the fitting procedure can be found in Ref.~\cite{Chatrchyan:2011hk}.

\begin{figure}[thb]
 \begin{center}
   \includegraphics[width=0.4\textwidth]{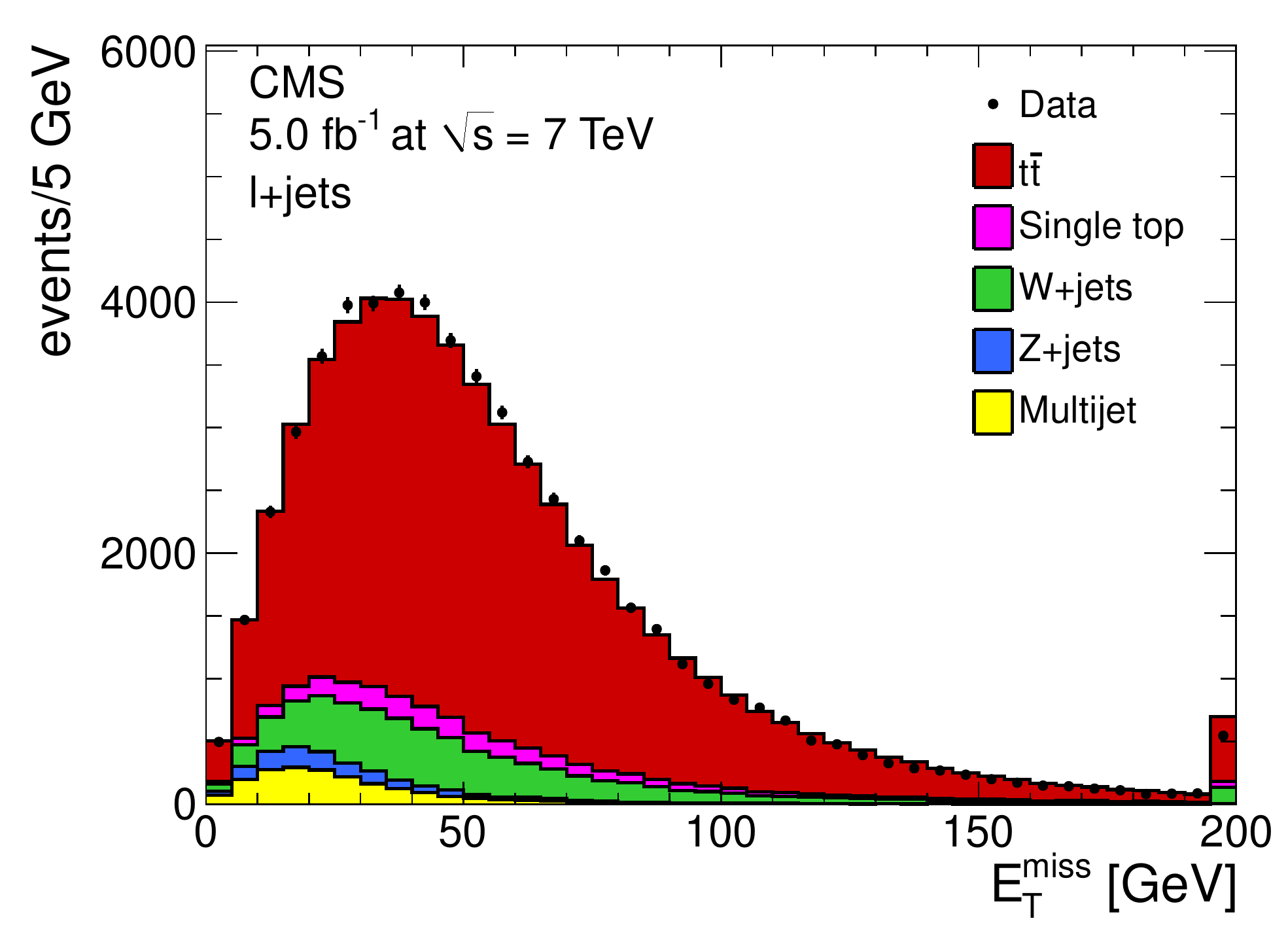}
   \includegraphics[width=0.4\textwidth]{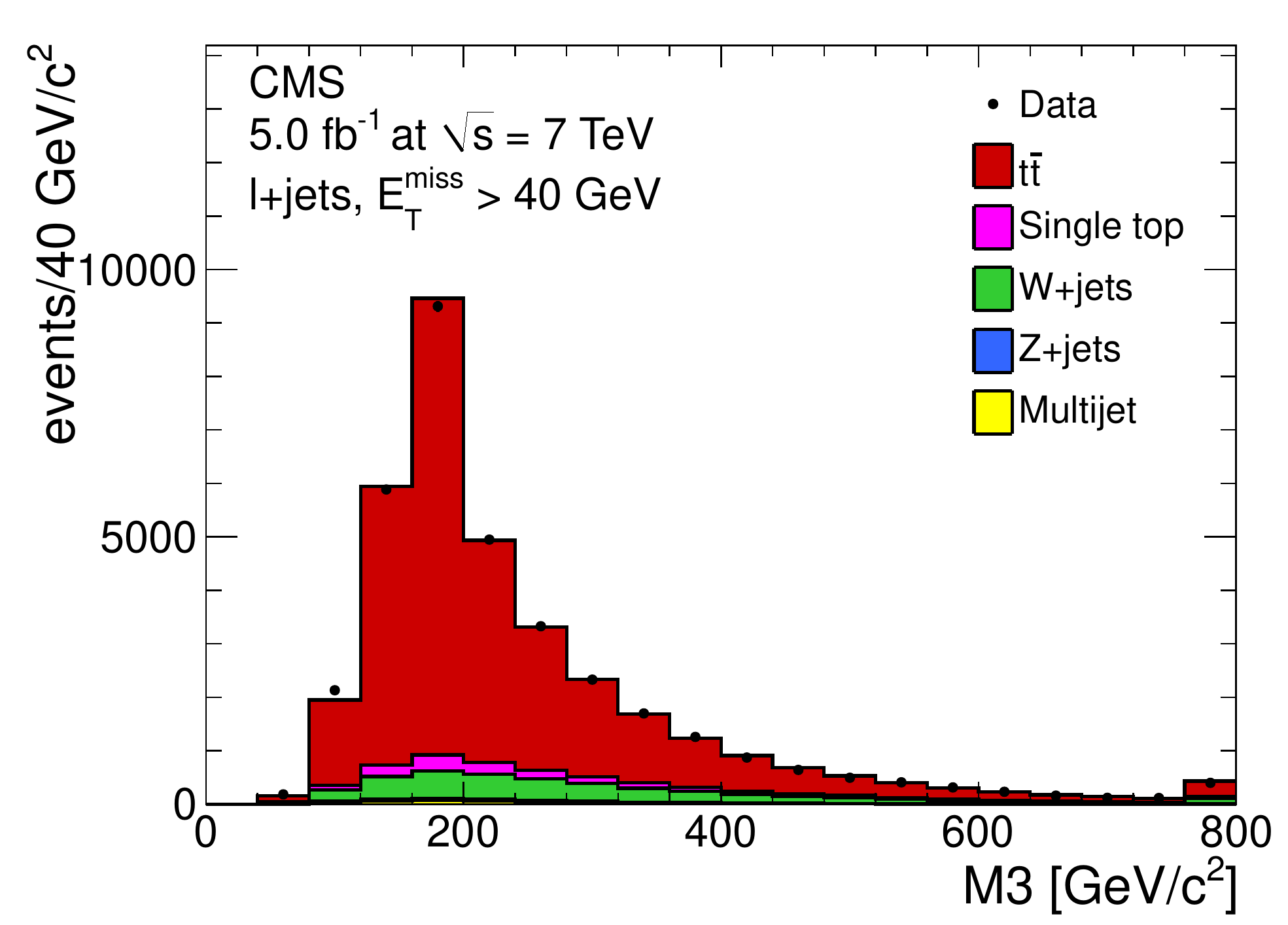}
       \caption{Comparison of the combined lepton$+$jets data with simulated contributions for the distributions in $\ETmiss$ (\cmsLeft) and M3 (\cmsRight). The simulated signal and background contributions are normalized to the results of the fits given in Table~\ref{tab:FitResults}.}
 \label{fig:DataMCfitresult}
\end{center}
\end{figure}

The resulting rates for the different processes can be correlated with each other, which has to be propagated to the calculation of the statistical uncertainty of the measured \ttbar charge asymmetry. The largest correlations are found between the rates of the \cPZ$+$jets and multijet backgrounds ($-$20\%) and between the rates for the \PW$^{+}+$jets and \PW$^{-}+$jets backgrounds ($+$12\%). All other correlations among the fit parameters are found to be small. Table~\ref{tab:FitResults} summarizes the results of the fits, along with their uncertainties. Figure~\ref{fig:DataMCfitresult} shows the measured \ETmiss and M3 distributions, with the individual simulated contributions normalized to the results from the fit.

\begin{table*}[htb]
 \topcaption{\label{tab:FitResults}Results for the numbers of events for background (BG) and $\ttbar$ contributions from fits to data, along with their uncertainties. The quoted uncertainties are of statistical nature with the exception of the uncertainties on the numbers for the single top and Z$+$jets backgrounds, which reflect the widths of the Gaussian rate constraints used in the likelihood fit (see text).}
 \begin{center}
    \begin{tabular}{lrrr} \hline
      Process                          & Electron$+$jets     & Muon$+$jets~~             &  Total~~~~~~~\\ \hline
      Single top ($t$ $+$ tW)          & $ 1113\pm 338$~~    &    $ 1418 \pm 505 $~~     & $ 2532 \pm 608 $~~\\
      \PWp$+$jets                  & $ 1818 \pm 227 $~~  &    $ 1807 \pm 290 $~~     & $ 3625 \pm 369 $~~\\
      \PWm$+$jets                  & $ 1454 \pm 224 $~~  &    $ 1320 \pm 275 $~~     & $ 2773 \pm 355 $~~\\
      \cPZ$+$jets                      & $ 535 \pm 153 $~~   &    $ 600 \pm 170 $~~      & $ 1135 \pm 229 $~~\\
      Multijet                         & $ 1142 \pm 227 $~~  &     $ 863 \pm 209 $~~     & $ 2005 \pm 308 $~~\\ \hline
      Total  BG                        & $ 6062 \pm 540 $~~  &    $ 6008 \pm 698 $~~     & $ 12070 \pm 882 $~~\\
      \ttbar                           & $ 18634 \pm 390 $~~ &    $ 26976  \pm 468 $~~   & $ 45610 \pm 609 $~~\\  \hline
      Observed data                    &     24705 ~~~~~~    &   32992   ~~~~~~          &      57697~~~~~~ \\ \hline
 \end{tabular}
\end{center}
\end{table*}

\section{Measurement of the \texorpdfstring{\ttbar}{t-tbar} charge asymmetry}

The measurement of the \ttbar charge asymmetry is based on the fully reconstructed four-momenta of the top quarks and antiquarks in each event. We reconstruct the leptonically decaying \PW~boson from the measured charged lepton and $\ETmiss$, and associate the measured jets in the event with the quarks in the \ttbar decay chain. The reconstruction procedure is described in detail in Ref.~\cite{Chatrchyan:2011hk}.

\begin{figure}[thb]
 \begin{center}
     \includegraphics[width=0.4\textwidth]{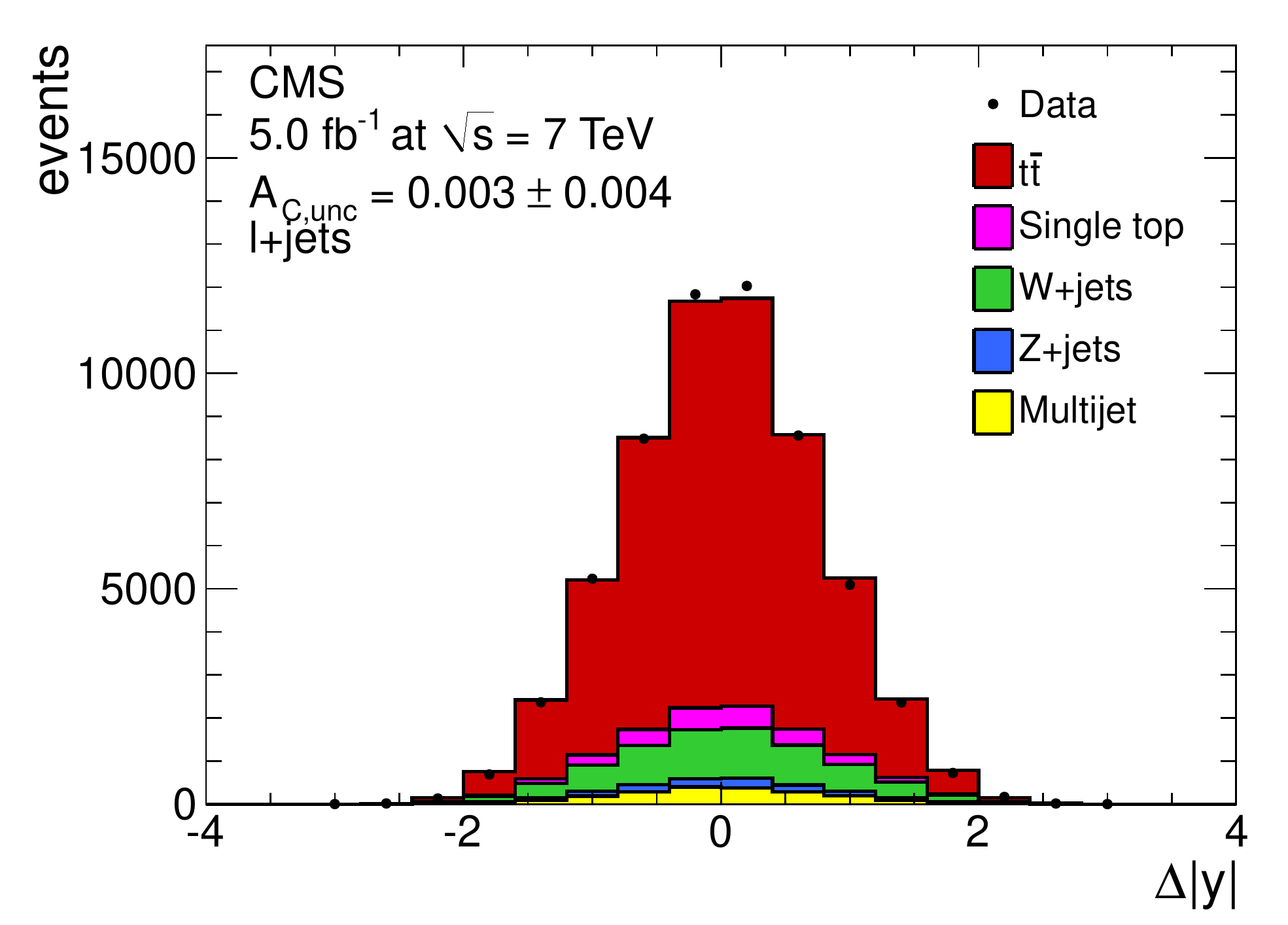}
     \caption{Comparison of the combined lepton$+$jets data with simulated contributions for the distributions in $\Delta|y|$. The simulated signal and background contributions are normalized to the results of the fits given in Table~\ref{tab:FitResults}.}
 \label{fig:DataMCfitresult2}
 \end{center}
\end{figure}

The reconstructed top quark and antiquark four-momenta are used to obtain the inclusive (see Fig.~\ref{fig:DataMCfitresult2}) and differential distributions of $\Delta|y|$ and the charge asymmetry is calculated from the number of entries with $\Delta|y| > 0$ and $\Delta|y| < 0$. In case of the differential measurements, the asymmetries are calculated separately for the different bins in the kinematic variable $V_{i}$, where $V_{i}$ is either $|y_{\ttbar}|$, $p_{\mathrm{T}}^{\ttbar}$, or $m_{\ttbar}$. To allow for a comparison of the resulting asymmetry and the predictions from theory, the reconstructed distributions of $\Delta|y|$ and the three kinematic variables have to be corrected for background contributions, reconstruction effects, and selection efficiencies.

In the first correction step, the distributions of background processes are normalized to the estimated rates (see Table~\ref{tab:FitResults}) and subtracted from the data, assuming Gaussian uncertainties on the background rates as well as on statistical fluctuations in the background templates. The correlations among the individual background rates are taken into account.

The background-subtracted distributions are transformed from the reconstruction level to the particle level after event selection, and from there to the particle level before event selection. The corrections are achieved by applying a regularized unfolding procedure to the data~\cite{Blobel:2002pu} through a generalized matrix-inversion method. In this method, the perturbing effects are described by a smearing matrix $S$ that translates the true spectrum $\vec{x}$ into the measured spectrum $\vec{w} = S\vec{x}$. As reconstruction and selection effects factorize, the smearing matrix $S$ can be constructed as the product of a migration matrix and a diagonal matrix with the efficiencies for each of the bins on the diagonal, and all other elements set to zero. The unfolding procedure used for the inclusive measurement --- described in detail in Ref.~\cite{Chatrchyan:2011hk} --- can be generalized to deal also with two-dimensional distributions.

\begin{figure*}[tbh]
 \begin{center}
   \includegraphics[width=0.4\textwidth]{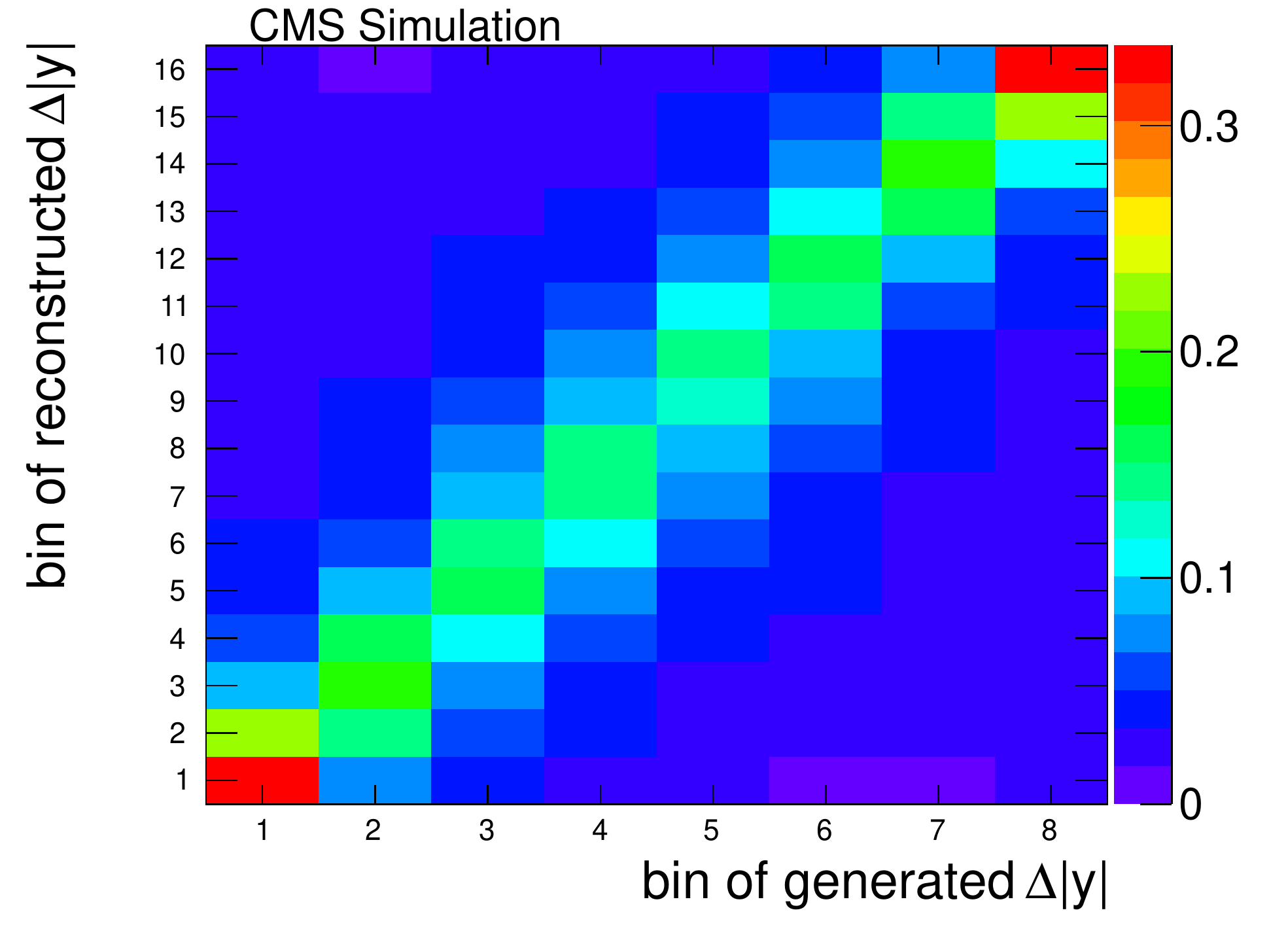}
   \includegraphics[width=0.4\textwidth]{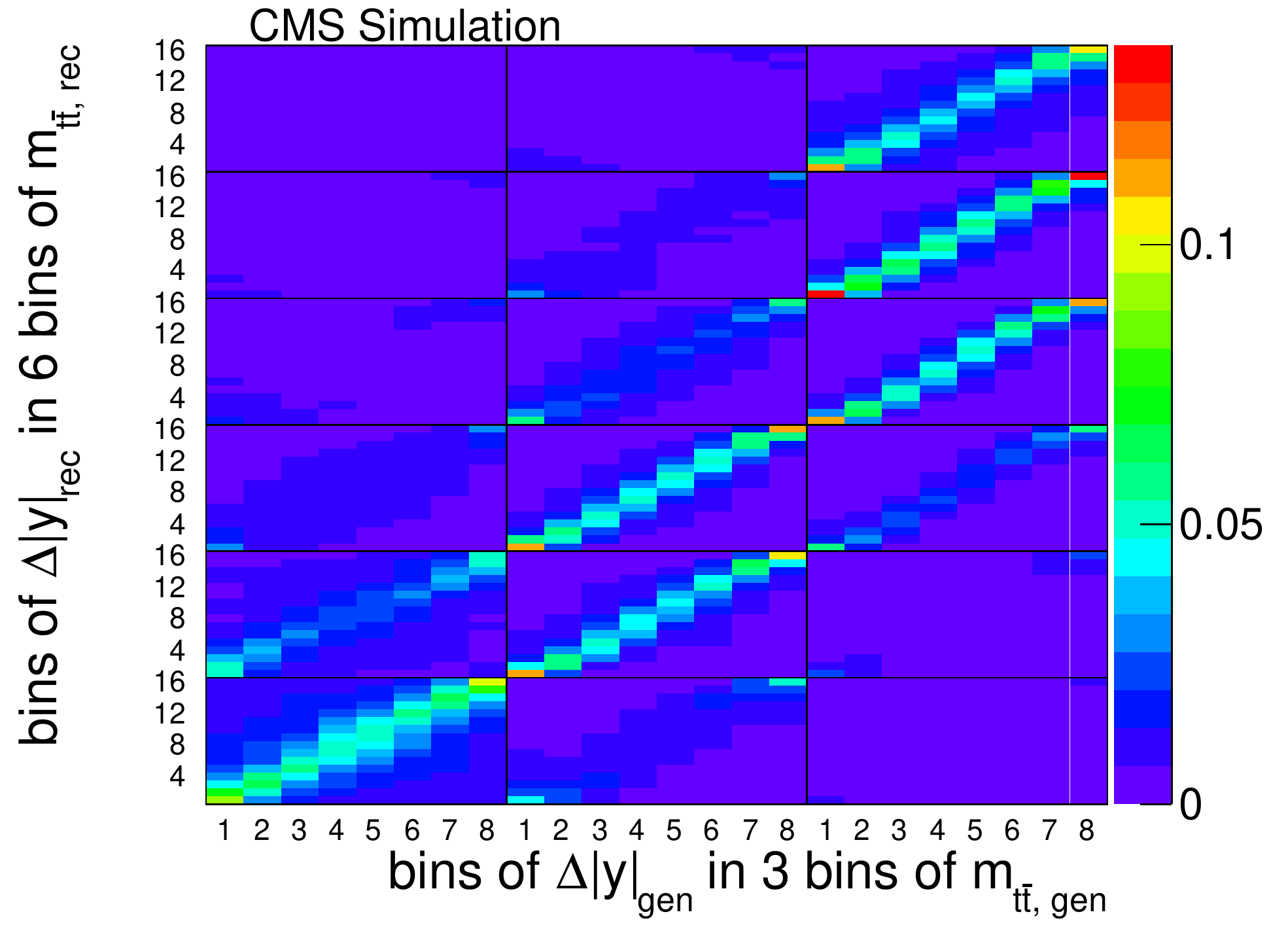}
   \includegraphics[width=0.4\textwidth]{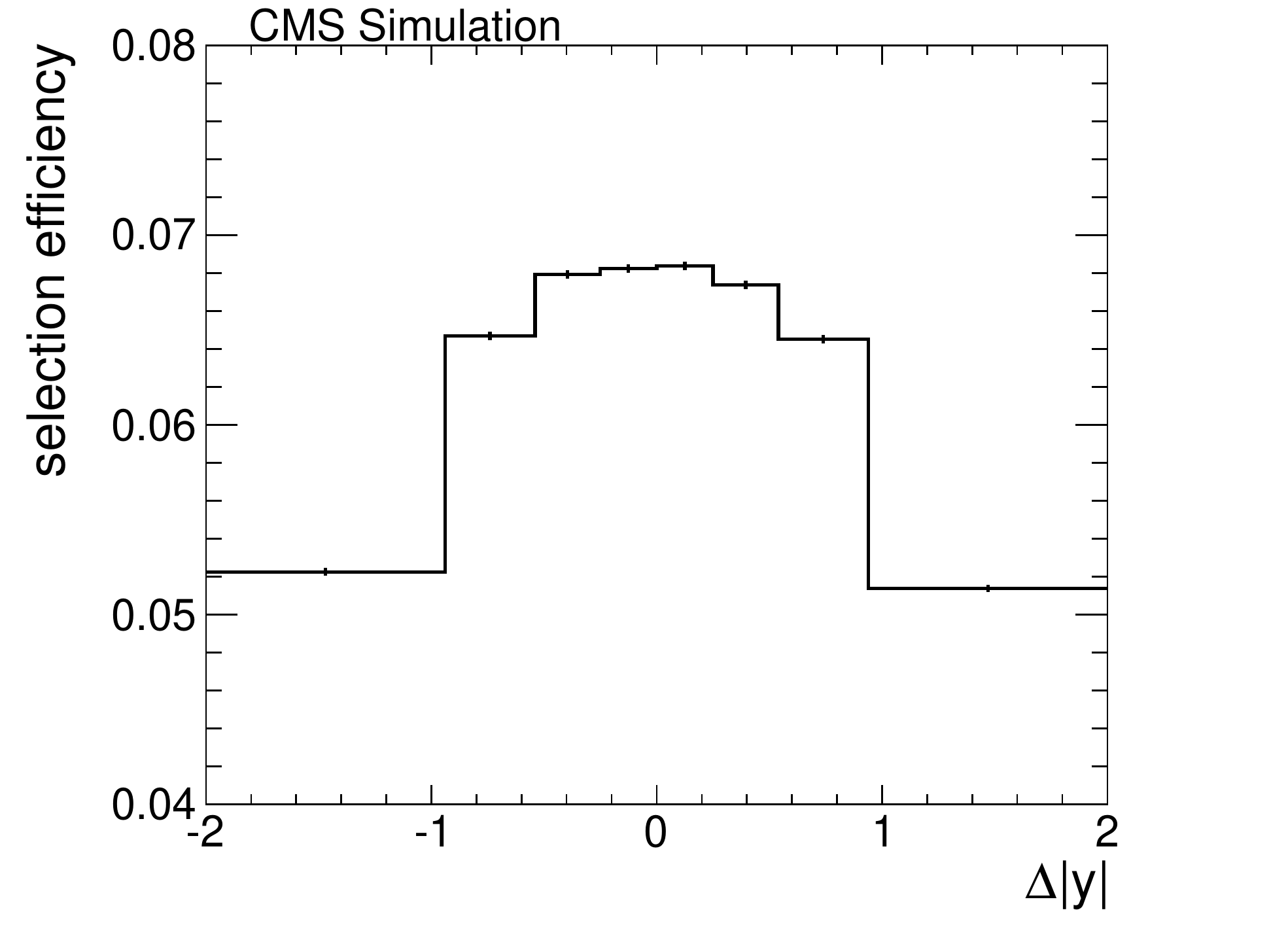}
   \includegraphics[width=0.4\textwidth]{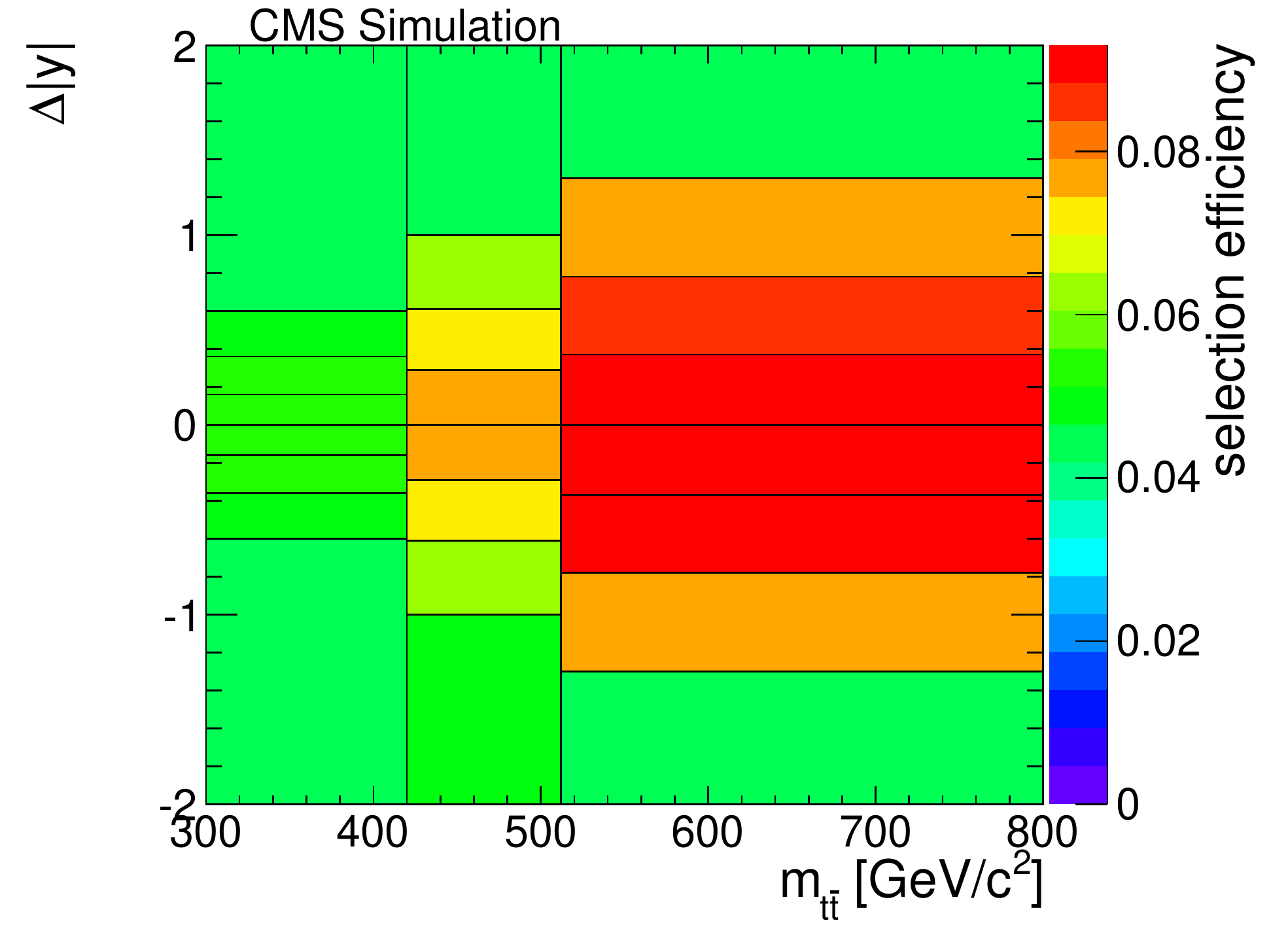}
    \caption{Migration matrix (upper row) between the generated and the reconstructed values after the event selection for $\Delta|y|$ (left) and for the measurement differential in $m_{\ttbar}$ (right). Selection efficiency (lower row) as a function of generated $\Delta|y|$, defined with respect to inclusive \ttbar production (left) and for the measurement differential in $m_{\ttbar}$ (right). }
 \label{fig:Asy_MigrationMatrices}
 \end{center}
\end{figure*}

The binning choice for a two-dimensional unfolding procedure has to fulfill some requirements in order to stabilize the unfolding procedure and to avoid a loss of resolution. For the applied unfolding procedure it is advised to use twice as many bins for the reconstructed spectra as for the unfolded spectra~\cite{Blobel:2002pu}. We use 16 (8) bins for the reconstructed (unfolded) $\Delta |y|$ distribution and 6 (3) bins in the reconstructed (unfolded) $V_{i}$ distributions. Furthermore it is desirable that the number of entries in each bin of the reconstructed distributions as well as in the unfolded distributions be approximately equal. The ranges for the bins in the unfolded kinematic variables are [0 -- 0.41; 0.41 -- 0.90; 0.90 -- $\infty$] for $|y_{\ttbar}|$, [ 0 -- 23; 23 -- 58; 58 -- $\infty$] for $p_{\mathrm{T}}^{\ttbar}$ in \GeVc, and [0 -- 420; 420 -- 512; 512 -- $\infty$] for $m_{\ttbar}$ in \GeVcc. The binning choice for $\Delta |y|$ is different in each bin of $V_{i}$, resulting in different amounts of vertical overlap between horizontally neighbouring bins in the two-dimensional distributions (for illustration see the binning in Fig.~\ref{fig:Asy_MigrationMatrices}, lower right). For the regularization of these distributions used in the differential measurements all combinations of neighbouring bins are considered. Due to the partial vertical overlap of horizontally neighbouring bins for a given central bin there are up to four possible combinations, each  weighted with a factor considering the amount of vertical overlap.

We use  separate migration matrices and selection efficiencies for the inclusive measurement and the three differential measurements, obtained from \ttbar events simulated with \POWHEG. Figure~\ref{fig:Asy_MigrationMatrices} shows the migration matrices for the inclusive measurement and for the differential measurement in $m_{\ttbar}$, as an example for the three migration matrices for the differential measurements. While for the inclusive measurement the migration matrix describes the migration of selected events from true values of $\Delta|y|$ to different reconstructed values, for the migration matrices of the differential measurements not only the migration between bins of $\Delta|y|$ has to be taken into account, but also the migration between bins of $V_{i}$. The migration matrices for the differential measurements feature on large scale a grid of $6 \times 3$ bins in $V_{i}$ with each of these bins hosting a $16 \times 8$ migration matrix describing the migration between different $\Delta|y|$ values. The values of $\Delta|y|$ and $V_{i}$ affect the probability for an event to survive the event selection criteria. The selection efficiencies as a function of $\Delta|y|$ for the inclusive measurement and as a function of $\Delta|y|$ and $m_{\ttbar}$ for the differential measurement in $m_{\ttbar}$ are depicted in Fig.~\ref{fig:Asy_MigrationMatrices}. The nearly symmetric shapes of the efficiency distributions imply that the effect of the event selection criteria on the \ttbar charge asymmetry is small.

The performance of the unfolding algorithm is tested in sets of pseudoexperiments, each of which provides a randomly generated sample distribution. For each pseudoexperiment the number of events from each contributing process is determined from a Poisson distribution around the mean of a Gaussian distribution centred around the measured event rate given in Table~\ref{tab:FitResults}, with a width corresponding to the respective uncertainty. We randomly draw the resulting number of events for each process from the respective simulated sample to generate distributions for each pseudoexperiment. Each distribution is then subjected to the background subtraction and unfolding procedure described above.

The average asymmetries from 50\,000 pseudoexperiments for the inclusive as well as for the differential measurements agree well with the true asymmetries in the sample used to model the signal component and the pull distributions agree with expectations, indicating that the treatment of uncertainties is consistent with Gaussian behavior. To test the unfolding procedure for different asymmetries, we reweight the events of the default \ttbar sample according to their $\Delta|y|$ value with a factor $w = k \cdot \Delta|y| + 1$, to artificially introduce asymmetries between $-0.2$ and $+0.2$, and then perform  pseudoexperiments for each of the reweighted distributions. For the differential measurements this test is performed in each of the three bins of $V_{i}$ separately. In all cases we find only negligible, if any deviations of the ensemble means from the input values for the asymmetry. In addition to this global reweighting of events, one can define the reweighting factor $w$ as a function of one of the kinematic variables, $w = k(V_{i})\cdot\Delta|y| + 1$. Four scenarios with $k$ rising or falling linearly with $V_i$ and one scenario in which $k$ rises quadratically are tested, generating asymmetries between $-0.1$ and $+0.1$. The effect of this reweighting dependent on $V_i$ is tested in all three possibilities to measure $A_C$ as a function of $V_j$. These scenarios serve as tests of the model-independence of the unfolding procedure, and observed deviations from the expectations are considered for the estimate of the systematic uncertainties of the measurement.

\section{Estimation of systematic uncertainties}

Systematic uncertainties with an impact on the differential selection efficiency, on the reconstructed top-quark momenta, or on the background rates can bias the results. To evaluate each source of systematic uncertainty, we repeat the background estimation and the measurement of $A_C$ using modified simulated samples. The expected systematic uncertainty for each source is taken to be the shift in the values of the corrected asymmetry between the default measurement and the one using the modified templates. The systematic uncertainties can be divided into three different categories: experimental sources, uncertainties in the modeling of the signal and background processes, and uncertainties due to the applied unfolding procedure.

The following experimental sources of systematic uncertainties are evaluated: variations in the jet energy scale (JES), jet energy resolution (JER), b-tagging efficiency, and the lepton selection efficiency. In order to derive the modified templates the corrections on JES and JER for simulated events are changed by $\pm 1$ standard deviations of their $\eta$- and \pt-dependent uncertainties. The overall scaling factor of the b-tagging efficiency does not affect the measurement, only $\eta$-dependent variations could in principle change the results. We therefore reweight events with b-tagged jets in the central region and in the forward regions maximally different within the b-tagging efficiency uncertainty. The effect on the $A_C$ measurement is found to be negligible. In a similar manner, we vary the scale factor for the lepton selection efficiency within its uncertainties ($\pm 1\%$ for muons; $\pm 2\%$ for electrons), this time with maximally different weighting factors for positively and negatively charged leptons, as a possible difference could lead to artificial asymmetries. This conservative treatment covers possible detector asymmetries as well as the probability of mismeasuring the lepton charge.

Regarding the simulation of signal and background processes, several sources of systematic uncertainties are evaluated. The uncertainty associated with the choice of the event generator used for modeling the \ttbar signal is estimated by using simulated events generated with \MADGRAPH instead of \POWHEG for the determination of the smearing matrix. In addition, a signal sample with a different hadronization and shower modeling has been used (\MCATNLO~4.0~\cite{Frixione:2002ik} interfaced to \HERWIG~\cite{corcella-2002}) to estimate the systematic uncertainty due to this part of the event generation. The effects of variations in the factorization and renormalization scales ($Q^{2}$) are estimated for \PW$+$jets and \ttbar events. For this purpose the strong coupling constant $\alpha_s$ and the parton distribution functions (PDF) are recalculated for each event for the varied $Q^{2}$ scale --- either multiplied with a factor of 4 or 0.25. The $Q^{2}$ scale is varied independently for W$+$jets and \ttbar processes and the estimated uncertainties have been added in quadrature to obtain the resulting systematic uncertainty on the measurement. The systematic uncertainties on the measured asymmetry from the choice of PDFs for the colliding protons used in the simulated events are estimated using the CTEQ6.6~\cite{cteq66} PDF set and the LHAPDF~\cite{lhapdf} package. In addition, the effect of variations in the frequency of occurrence of pileup events, overlaid on the simulated signal and background events, is estimated.

As the first step of the correction of the measured distributions is the subtraction of the background, the measurement is sensitive to the asymmetries present in the background model and we therefore evaluate the influence of possible mismodeling of the two backgrounds which are most sensitive to mismodeling. The rates for positively charged and negatively charged \PW~bosons are asymmetric and since the distributions of the two processes are slightly different, a mismodeling could artificially produce asymmetries. To estimate the effects from possible mismodeling of the \PW$+$jets background component, the templates for \PW$^{+}$ and \PW$^{-}$ processes are interchanged. Further studies are performed using a control sample enriched in \PW$+$jets events by selecting only events without b-tagged jets. The small observed differences between simulation and data in the inclusive distributions, as well as in the differential ones, are well encapsulated by the applied method to vary the \PW$+$jets template. The other background process that can show artificial asymmetries is the multijet background. This is the case if the rates for positively and negatively charged leptons differ in this sample. The multijet background is modeled using events from a sideband region, defined by inverting the requirements on the isolation of the charged lepton candidates. In these events the lepton rapidity is on average larger than the jet rapidities. As a result, the reconstructed leptonically decaying top quark candidates have on average a larger absolute value of the rapidity than the hadronically decaying top quark candidates, which in the end leads to different mean values of $\Delta|y|$ for events with positively and negatively charged leptons, respectively. To account for this effect, we invert the sign of $\Delta|y|$ for each event and use this altered template to model the multijet background.

\begin{table}[tbh]
 \begin{center}
\topcaption{\label{tab:incSyst}
Systematic uncertainties for the inclusive measurement of $A_{C}$.}
  \begin{tabular}{r|c} \hline
     Systematic uncertainty    & Shift ($\pm$) in inclusive $A_{C}$  \\ \hline
     JES                       & $0.003$   \\
     JER                       & $0.002$   \\
     Lepton ID/sel. efficiency & $0.006$  \\
     Generator                 & $0.001$   \\
     Hadronization             & $0.001$  \\
     $Q^2$ scale               & $0.002$  \\
     PDF                       & $0.002$  \\
     Pileup                    & $< 0.001$   \\
     \PW$+$jets                & $0.004$  \\
     Multijet                  & $0.001$  \\
     Migration matrix          & $0.002$   \\
     Model dependence          & $0.007$  \\ \hline
     Total                     & $0.011$ \\\hline
 \end{tabular}
\end{center}
\end{table}

The third category of systematic uncertainties deals with the impact of the limited number of simulated events and possible violations of the assumption that the applied unfolding procedure is model-independent. The impact of statistical uncertainties of the entries in the migration matrices is evaluated by repeating the measurement with altered migration matrices, where each element is varied within its statistical uncertainties. In addition to these uncertainties, we estimate the influence of possible dependencies of the asymmetry on one of the three kinematic variables $V_{i}$ (``model-dependence''). We perform pseudoexperiments with reweighted simulated signal samples and evaluate the differences between true and measured asymmetries in various reweighting scenarios, as described above. We take the average of the absolute values of the observed deviations and assign it as systematic uncertainty.

The contributions of the different sources of systematic uncertainties to the total uncertainty of the inclusive measurement are summarized in Table~\ref{tab:incSyst}. The total systematic uncertainty is smaller than the one obtained in Ref.~\cite{Chatrchyan:2011hk}. The two main changes in the evaluation of systematic uncertainties with respect to Ref.~\cite{Chatrchyan:2011hk}, which account for this difference, are discussed below. Variations in the threshold for the matching of matrix elements and parton shower evolution for the simulation of the $\ttbar$ signal~\cite{MLM}, causing the largest contribution to the total systematic uncertainty in the previous measurement, have no impact on the present measurement due to the usage of the NLO event generator \POWHEG for modeling the \ttbar signal. Furthermore, we do not quote a separate uncertainty due to variations in the amount of ISR and FSR on the measurement, as this contribution is covered by the uncertainties due to the choice of the $Q^{2}$ scale and the model-dependence systematic. The probability for additional radiation increases with decreasing $Q^{2}$ and vice versa. Due to the strong correlation between the amount of additional radiation and the transverse momentum of the \ttbar system, the variation of the generated asymmetry as a function of $p_{\mathrm{T}}^{\ttbar}$, as done in the estimation of the model dependence uncertainty, is also suited to estimate the effects of variations in the amount of ISR/FSR on the measurement.

The systematic uncertainties on the differential measurements are included in the error bars of the corrected differential distributions (see Fig.~\ref{fig:results}). Depending on $V_{i}$ and the actual bin, the contributions from the different sources vary. The largest contributions arise from variations in the JES and the conservatively estimated uncertainties due to lepton selection efficiency and model-dependence as well as the statistical fluctuations of the migration matrix. The generator and hadronization uncertainty play a significant role for the measurements differential in $m_{\ttbar}$ and $p_{\mathrm{T}}^{\ttbar}$. The modeling of the major background, the \PW$+$jets process, is significant in the third bin of $m_{\ttbar}$ and $|y_{\ttbar}|$.

\section{Results}

\begin{figure*}[h!tb]
 \begin{center}
   \includegraphics[width=0.4\textwidth]{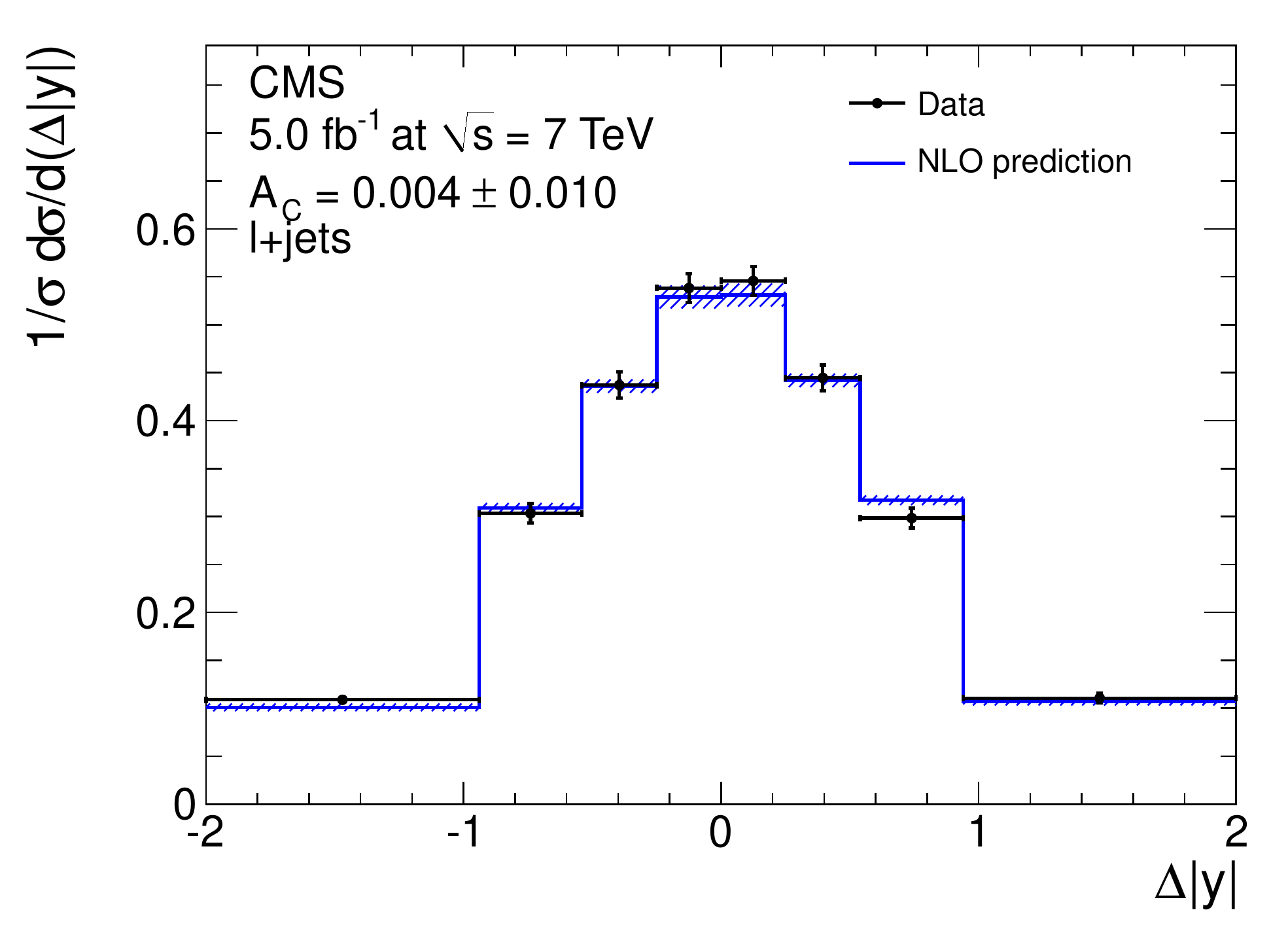}
   \includegraphics[width=0.4\textwidth]{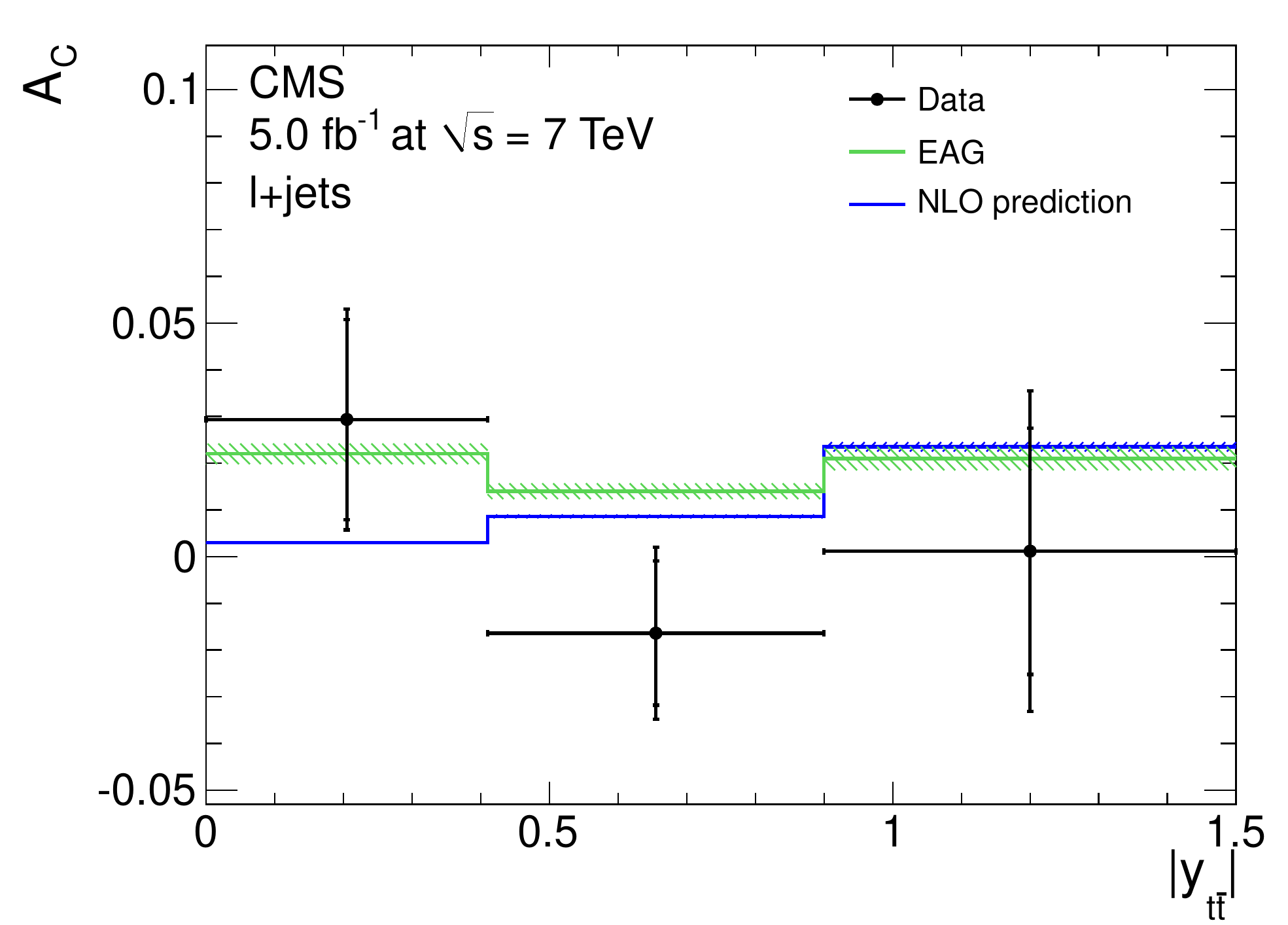}
   \includegraphics[width=0.4\textwidth]{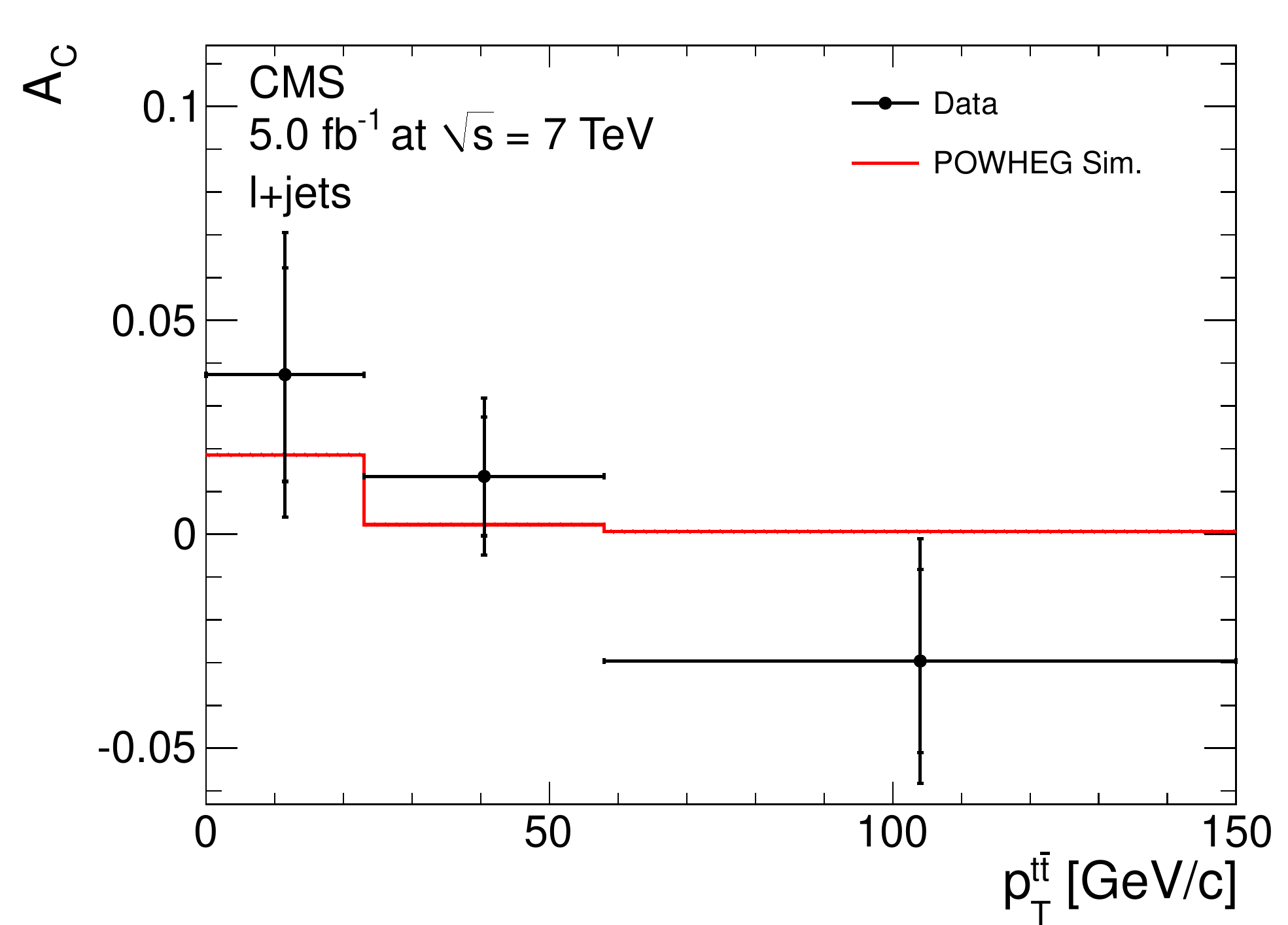}
   \includegraphics[width=0.4\textwidth]{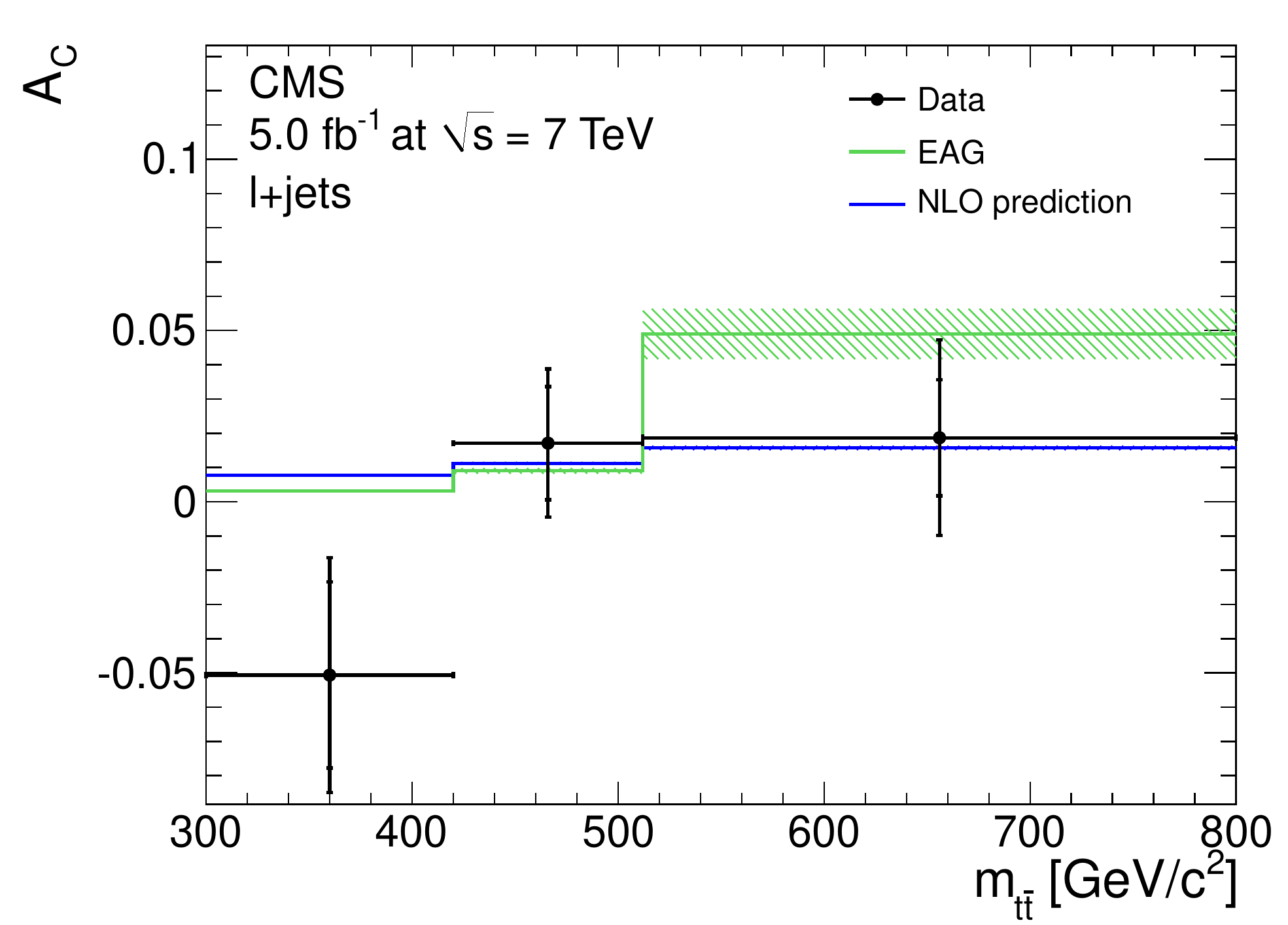}
   \caption{Unfolded inclusive $\Delta|y|$ distribution (upper left), corrected asymmetry as a function of $|y_{\ttbar}|$ (upper right), $p_{\mathrm{T}}^{\ttbar}$ (lower left), and $m_{\ttbar}$ (lower right). The measured values are compared to NLO calculations for the SM --- based on the calculations of Ref.~\cite{Kuhn:2011ri} --- and to the predictions of a model featuring an effective axial-vector coupling of the gluon (EAG)~\cite{Brooijmans:2012yi}. The error bars on the differential asymmetry values indicate the statistical and total uncertainties, determined by adding statistical and systematic uncertainties in quadrature. The shaded areas indicate the theoretical uncertainties on the NLO calculations.}
 \label{fig:results}
 \end{center}
\end{figure*}

The unfolded $\Delta|y|$ distribution, shown in Fig.~\ref{fig:results}, is used to calculate the corrected inclusive asymmetry:

\begin{equation}
 A_{C} = 0.004 \pm 0.010 ~(\text{stat.})\,\pm 0.011 ~(\text{syst.})~.
\end{equation}

Table~\ref{tab:Results_Inc} gives the values of the measured inclusive asymmetry at the different stages of the analysis.

\begin{table*}[hbt]
\renewcommand{\arraystretch}{1.2}
\topcaption{\label{tab:Results_Inc} The measured inclusive asymmetry at the different stages of the analysis and the corresponding theoretical prediction from the SM.}
 \begin{center}
    \begin{tabular}{lr} \hline
Uncorrected               & $0.003 \pm 0.004 ~(\text{stat.}) $\\
BG-subtracted             &$ 0.002 \pm 0.005 ~(\text{stat.})\,\pm 0.003 ~(\text{syst.}) $\\
Final corrected           &$ 0.004 \pm 0.010 ~(\text{stat.})\,\pm 0.011 ~(\text{syst.})$\\ \hline
Theoretical prediction  (SM)   &$0.0115 \pm 0.0006 $  \\ \hline
 \end{tabular}
\end{center}
\end{table*}

The results of the three differential measurements can be found in Table~\ref{tab:Results_Diff} and Fig.~\ref{fig:results}. The measured values are compared to the SM predictions --- based on the calculation of Ref.~\cite{Kuhn:2011ri} --- and as an illustrative example to the predictions from a BSM model that introduces an anomalous effective axial-vector coupling to the gluon at the one-loop level~\cite{PhysRevD.85.074021, Brooijmans:2012yi}. The gluon-quark vertex is treated in the approximation of an effective field theory with an order of 1\TeV scale for new physics contributions. This is a model that can explain the strong dependence of the forward-backward asymmetry on $m_{\ttbar}$ as seen by CDF. As the theoretical predictions are normalized to the leading-order cross section and $p_{\mathrm{T}}^{\ttbar}$ is zero at LO, no theoretical predictions are available for this differential measurement. Instead, we compare the measured asymmetries with the predictions obtained from \POWHEG simulation. Within the uncertainties the data do not show any significant asymmetry and all measured values are consistent with a null asymmetry as well as with the SM predicted values. The current level of precision is not yet sufficient to discriminate the explored BSM model either.

\begin{table*}[hbt]
\renewcommand{\arraystretch}{1.2}
\topcaption{\label{tab:Results_Diff}The corrected asymmetry values in three bins of the kinematic variables $|y_{\mathrm{t\bar{t}}}|$, $p_{\mathrm{T, t\bar{t}}}$, and $m_{\mathrm{t\bar{t}}}$ with statistical and systematic uncertainties, along with the SM predictions (in case of $p_{\mathrm{T}}^{\ttbar}$ we compare to the values obtained from \POWHEG simulation).}
 \begin{center}
    \begin{tabular}{lccc} \hline
  Kinematic variable     &    $A_{C}$ in bin 1                &  $A_{C}$ in bin 2                     & $A_{C}$ in bin 3 \\ \hline
$|y_{\ttbar}|$           &$     0.029 \pm   0.021\,\pm 0.010 $&$  -0.016  \pm  0.015\,\pm 0.010 $     & $  0.001 \pm  0.026\,\pm 0.022 $ \\
$|y_{\ttbar}|$(SM pred.)   &  $   0.0030  \pm  0.0002   $       &  $ 0.0086      \pm  0.0004    $       &   $  0.0235   \pm   0.0010  $    \\ \hline
$p_{\mathrm{T}}^{\ttbar}$  &$     0.037  \pm  0.025\,\pm 0.022 $& $   0.014 \pm 0.014\,\pm 0.012   $  & $  -0.030 \pm    0.021\,\pm0.019  $\\
$p_{\mathrm{T}}^{\ttbar}$ (simulation) & $0.0185 \pm 0.0004$    & $ 0.0022 \pm  0.0004$               & $ 0.0006\pm 0.0004$ \\ \hline
$m_{\ttbar}$             & $   -0.051   \pm 0.027\,\pm 0.021 $&  $ 0.017  \pm    0.017 \,\pm 0.014 $  & $   0.019 \pm  0.017 \,\pm 0.023 $ \\
$m_{\ttbar}$ (SM pred.)    & $     0.0077 \pm  0.0003 $         & $ 0.0112 \pm 0.0004   $  & $ 0.0157 \pm 0.0006     $ \\\hline
 \end{tabular}
\end{center}
\end{table*}

\section{Conclusion}

An inclusive and three differential measurements of the charge asymmetry in \ttbar production at the LHC have been presented. Events with top-quark pairs decaying in the electron$+$jets and muon$+$jets channels were selected and a full \ttbar event reconstruction was performed to determine the four-momenta of the top quarks and antiquarks. The observed distributions were then corrected for acceptance and reconstruction effects. Although the measured values constitute the most precise determination of the \ttbar charge asymmetry at the LHC to date, the current precision does not yet allow distinguishing a zero asymmetry from the values predicted in the standard model or in BSM theories. The reported results nonetheless indicate that LHC data disfavor large deviations from the SM predictions. To get a quantitative picture and to answer the question whether or not the observed slight difference between $A_{FB}$ at the Tevatron and $A_C$ at the LHC is due to BSM physics, it is essential to further explore $A_C$. This is especially true in kinematic regions where the $\cPq\cPaq \rightarrow \ttbar$ contribution, and thus the charge asymmetry, is enhanced.

\section*{Acknowledgments}

We thank G. Rodrigo, and J.H. K\"uhn for fruitful discussions and congratulate our colleagues in the CERN accelerator departments for the excellent performance of the LHC machine. We thank the technical and administrative staff at CERN and other CMS institutes, and acknowledge support from: FMSR (Austria); FNRS and FWO (Belgium); CNPq, CAPES, FAPERJ, and FAPESP (Brazil); MES (Bulgaria); CERN; CAS, MoST, and NSFC (China); COLCIENCIAS (Colombia); MSES (Croatia); RPF (Cyprus); MoER, SF0690030s09 and ERDF (Estonia); Academy of Finland, MEC, and HIP (Finland); CEA and CNRS/IN2P3 (France); BMBF, DFG, and HGF (Germany); GSRT (Greece); OTKA and NKTH (Hungary); DAE and DST (India); IPM (Iran); SFI (Ireland); INFN (Italy); NRF and WCU (Korea); LAS (Lithuania); CINVESTAV, CONACYT, SEP, and UASLP-FAI (Mexico); MSI (New Zealand); PAEC (Pakistan); MSHE and NSC (Poland); FCT (Portugal); JINR (Armenia, Belarus, Georgia, Ukraine, Uzbekistan); MON, RosAtom, RAS and RFBR (Russia); MSTD (Serbia); MICINN and CPAN (Spain); Swiss Funding Agencies (Switzerland); NSC (Taipei); TUBITAK and TAEK (Turkey); STFC (United Kingdom); DOE and NSF (USA).

Individuals have received support from the Marie-Curie programme and the European Research Council (European Union); the Leventis Foundation; the A. P. Sloan Foundation; the Alexander von Humboldt Foundation; the Belgian Federal Science Policy Office; the Fonds pour la Formation \`a la Recherche dans l'Industrie et dans l'Agriculture (FRIA-Belgium); the Agentschap voor Innovatie door Wetenschap en Technologie (IWT-Belgium); the Council of Science and Industrial Research, India; and the HOMING PLUS programme of Foundation for Polish Science, cofinanced from European Union, Regional Development Fund.

\bibliography{auto_generated}

\cleardoublepage \appendix\section{The CMS Collaboration \label{app:collab}}\begin{sloppypar}\hyphenpenalty=5000\widowpenalty=500\clubpenalty=5000\textbf{Yerevan Physics Institute,  Yerevan,  Armenia}\\*[0pt]
S.~Chatrchyan, V.~Khachatryan, A.M.~Sirunyan, A.~Tumasyan
\vskip\cmsinstskip
\textbf{Institut f\"{u}r Hochenergiephysik der OeAW,  Wien,  Austria}\\*[0pt]
W.~Adam, E.~Aguilo, T.~Bergauer, M.~Dragicevic, J.~Er\"{o}, C.~Fabjan\cmsAuthorMark{1}, M.~Friedl, R.~Fr\"{u}hwirth\cmsAuthorMark{1}, V.M.~Ghete, J.~Hammer, N.~H\"{o}rmann, J.~Hrubec, M.~Jeitler\cmsAuthorMark{1}, W.~Kiesenhofer, V.~Kn\"{u}nz, M.~Krammer\cmsAuthorMark{1}, D.~Liko, I.~Mikulec, M.~Pernicka$^{\textrm{\dag}}$, B.~Rahbaran, C.~Rohringer, H.~Rohringer, R.~Sch\"{o}fbeck, J.~Strauss, A.~Taurok, P.~Wagner, W.~Waltenberger, G.~Walzel, E.~Widl, C.-E.~Wulz\cmsAuthorMark{1}
% ...
\vskip\cmsinstskip
\dag:~Deceased\\
1:~~Also at Vienna University of Technology, Vienna, Austria\\
% ...

\end{sloppypar}
\end{document}